\documentclass[10pt,a4paper]{article}
\usepackage{graphicx}
\usepackage{amsmath}
\usepackage[small,bf]{caption}
\usepackage{booktabs}
\usepackage{subfig}
\usepackage[left=2.5cm,top=3cm,right=2.5cm,bottom=3cm]{geometry}

\usepackage{amsmath,amssymb}
\usepackage[affil-it]{authblk}
\usepackage{graphicx}
\usepackage[hidelinks]{hyperref}
\hypersetup{urlcolor=blue}
\begin{document}

\begin{center}
\Large\textbf{Lily Pad: Towards Real-time Interactive\\Computational Fluid Dynamics}\\[1em]
\large\textbf{Gabriel D. Weymouth},
University of Southampton, UK, \\
\href{mailto:G.D.Weymouth@soton.ac.uk}{\nolinkurl{G.D.Weymouth@soton.ac.uk}}\\[3em]
\end{center}

\thispagestyle{empty}
\pagestyle{empty}

\section{Introduction}

Amazing advances have been made in both computing resources and computational methodologies in the last 15 years - enabling first-principle predictions of ship science flows which are fast enough to be useful in engineering. In addition, open source projects such as OpenFOAM \cite{openfoam} provide a free and adaptable alternative to standard engineering software. However, despite the fact that computational fluid dynamics (CFD) software is now (relatively) fast and freely available, it is still amazingly difficult to use. CFD software is complex, fussy, opaque, and poorly designed \cite{Rouson2011}, even compared to other scientific software such as the free and open source R-project for statistical analysis \cite{r-project}.\footnote{To say nothing of software that people actually \textit{enjoy} using.} Inaccessible software imposes a significant entry barrier on students and junior engineers, and even senior researchers spend less time developing insights and more time on software issues \cite{Groen2015}.

Lily Pad \cite{lilypad} was developed as an initial attempt to address some of these problems. The goal of Lily Pad is to lower the barrier to CFD by adopting simple high-speed methods, utilizing modern programming features and environments, and giving immediate visual feed-back to the user. The resulting software focuses on the fluid dynamics instead of the computation, making it useful for both education and research.

It would be impossible to genuinely achieve this goal unless Lily Pad was genuine CFD solver. Therefore the full two-dimensional Navier-Stokes equations are solved and the exact body boundary conditions are applied. However, most of the complications plaguing CFD are avoided by using the Boundary Data Immersion Method (BDIM \cite{Weymouth2011JCP,Maertens2015}) to immerse solid bodies into the fluid domain. Despite this simplicity (or perhaps because of it), BDIM is very accurate, it has some nice analytic properties, and has been extensively validated.

\begin{figure}[t!]
	\centering
	\subfloat[]{
		\includegraphics[width=0.55\textwidth]{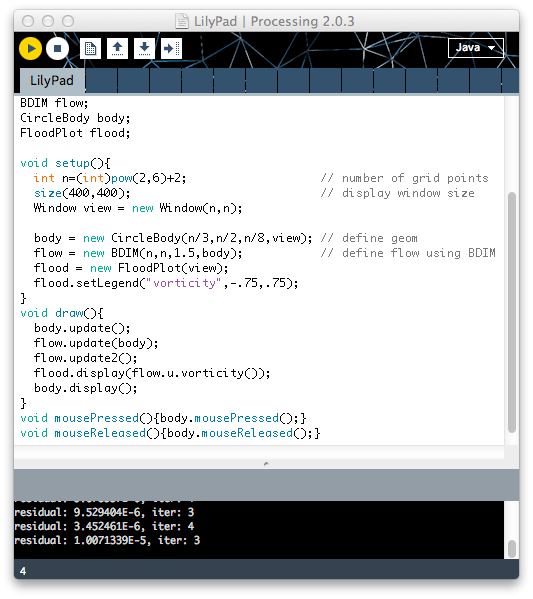}
		\label{fig:pde}}
	\subfloat[]{
		\includegraphics[width=0.4\textwidth]{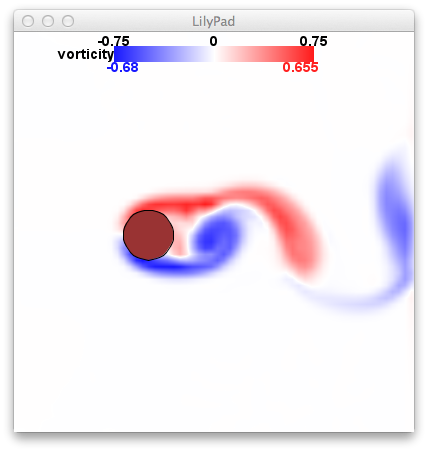}
		\label{fig:circle}}
	\caption{Lily Pad example in the Processing Development Environment: (a) complete example script, and (b) real-time interactive visualization.}
	\label{fig:default}
\end{figure}

Lily Pad is written in Processing \cite{processing}, a programming language initially based on Java which promotes ``software literacy within the visual arts and visual literacy within technology''. Lily Pad utilizes the Processing Development Environment, which integrates the writing, testing, and usage modes into a single platform. Specifically, a short script is written on the \textit{front-page} (Fig~\ref{fig:pde}), the code is executed by pressing a button, and the flow is simultaneously simulated and visualized in a pop-up window (Fig~\ref{fig:circle}). Fig~\ref{fig:default} is a complete working example, the default simulation of the flow around a circular cylinder. And it is interactive - the circle can be dragged around the window and the flow is updated in real-time.

While the default flow in Fig~\ref{fig:default} is at low resolution and features fairly simple physics, the method is completely extendable, and Lily Pad simulations have already been published in the Journal of Fluid Mechanics \cite{Weymouth2014JFM,Polet2015} and have been used to initiate and test many additional research projects, as summarized in the examples section. First, the numerical method and implementation in Lily Pad is summarized in more detail.

\section{Flow Solver Methodology}

Solving the incompressible Navier-Stokes equation numerically presents a few irreducible difficulties, namely time integration of the finite-scale nonlinear momentum equations and inversion of a Poisson equation for the pressure. However, the majority of the complexity in CFD software comes not from the fluid equations, but from coupling those equations to the irregular data at domain boundaries. The adoption of boundary-fitted meshes to achieve this coupling increases the memory requirements, computational effort and algorithmic complexity of the numerical method, as well as presenting conceptual and modeling difficulties to the user.

Lily Pad utilizes fast and robust methods for dealing with all of these issues, starting with the adoption of the Boundary Data Immersion Method for coupling the fluid and solid equations. The basic idea of BDIM (and all Immersed Boundary methods since \cite{Peskin1972}) is to adjust the equations of motion to account for the interaction of the fluid and solid and solve those equations on a simple numerical grid. In Lily Pad this is a uniform Cartesian grid. This avoids adopting questionable physical simplifications (such as treating fluids as particles, with issues documented in \cite{Kiara2013}) or hiding the numerical complexity in a black box (such as automatic mesh refinement). Instead, the analytic governing equations themselves are adjusted, allowing the numerical method to remain as simple and efficient as possible.

The full details are in \cite{Maertens2015}, but briefly, the analytic BDIM equations are obtained by the convolution of two governing equations, each valid over their own region of the total domain. In Lily Pad this is a body with prescribed velocity immersed in a 2D viscous fluid. Using a symmetric kernel with width $\epsilon$ for the convolution and keeping up to second order terms gives
\begin{eqnarray}
\vec{u}_\epsilon &=& \mu_0 \vec{f}+(1-\mu_0)\vec{b}+\mu_1\frac{\partial}{\partial n}\left(\vec{f}-\vec{b}\right) \label{eq:velo} \\
\vec\nabla\cdot\vec{u}_\epsilon &=& (1-\mu_0)\vec\nabla\cdot\vec{b}-\mu_1\frac{\partial}{\partial n}\left(\vec\nabla\cdot\vec{b}\right) \label{eq:pres}
\end{eqnarray}
where $\vec f$ and $\vec b$ are the update equations for the velocity in the fluid and body domains, and where $\mu_0$ and $\mu_1$ are the zeroth and first moments of the kernel function with respect to the fluid domain. The convoluted velocity $\vec u_\epsilon$ update equation~\eqref{eq:velo}, takes the form of a mixing equation, with the fluid equation used in the fluid domain (where $\mu_0=1,\ \mu_1=0$) the body equation used in the body domain (where $\mu_0=\mu_1=0$) and a combination of the two within $\epsilon$ of the boundary. The velocity divergence equation~\eqref{eq:pres} is used to solve for the pressure required to satisfy the divergence condition in both domains, which enables volume-changing bodies found in biological flow problems to be modeled. Three important points: (a) the normal derivative term in the velocity equation (active in the boundary region) enables accurate boundary layer simulation including separation prediction  \cite{Maertens2015}; (b) the presence of the kernel moments results in a variable coefficient Poisson equation for the pressure which implicitly enforces the correct pressure boundary conditions and significantly improves force prediction \cite{Weymouth2011JCP}; and (c) the body velocity equation $\vec b$ can be easily coupled to the state of the fluid, enabling prescribed or predicted fluid-structure interaction simulations \cite{Weymouth2014JFM}. BDIM is therefore capable of fluid-structure interaction predictions using a uniform Cartesian grid, and this is leveraged in Lily Pad to greatly simplify the software.

Lily Pad uses implicit Large Eddy Simulation (ILES, \cite{Margolin2013}) to model the scales below the finite grid spacing in order to avoid the standard turbulence model selection and parameterization issues. Depsite its simplicity (or, again, perhaps because of it) ILES has been shown to have greater effective subgrid resolution than explicit LES models \cite{Aspden2008}, but like all LES models it requires explicit time-stepping and is difficult to scale to high Reynolds number flows. Either a flux-limited QUICK or a Semi-Lagrangian convection scheme can be used. The Semi-Lagrangian scheme is universally stable, enabling larger time steps at the cost of reduced the accuracy near the body where the flow gradients are large. Additionally, because Lily Pad is two-dimensional, the results at intermediate Reynolds numbers will tend to have features of two-dimensional turbulence, which can be quite distinct from their three-dimensional counterparts \cite{Couder1986}. Within these limitations, the choice of ILES enables a wide variety of interesting flows to be simulated in Lily Pad quickly and easily.

The variable coefficient pressure equation is solved using a Multi-Grid method with Jacobi smoothing. The unit Cartesian grid, sparse symmetric matrix, and relatively small time steps make this extremely efficient; the residual is typically driven below the tolerance level in only a single V-cycle, making determination of the pressure \textit{less expensive} than evaluating the momentum equations.

\section{Implementation and Usage}

Lily Pad is written in Processing in an object-oriented programming style, using classes of objects to set up and solve the CFD system. The core classes are as follows:

\begin{itemize}
\item \texttt{Field} contains the data structures and method for handling scalar fields such as pressure. This includes methods for differentiation, interpolation\footnote{The unit background grid makes differentiation and interpolation operations nearly trivial.} and advection of the field, as well as setting boundary conditions and displaying the scalar values on-screen.

\item \texttt{VectorField} contains the methods for vector fields such as the fluid velocity, made up of two \texttt{Field} components. A projection operator (using Multi-Grid) is defined here in addition to vector differentiation and the convection/diffusion operators.

\item \texttt{Body} is the parent class of all the solid geometries which are used in Lily Pad. Methods to transform (translate, rotate), query (force?, moment?, distance?) and display the body are defined.

\item \texttt{BDIM} is the high-level class used on the \textit{front-page} to set-up and solve the general solid-fluid interaction problem described by equations \ref{eq:velo} and \ref{eq:pres}.
\end{itemize}

The default script in Fig~\ref{fig:default} shows how these classes and methods are used in Lily Pad. The Processing front-page is broken into four parts: variable declaration, the \texttt{setup} method, the \texttt{draw} method, and additional problem specific methods. In the default script three objects are declared:
\begin{verbatim}
  BDIM flow;
  CircleBody body;
  FloodPlot flood;
\end{verbatim}
where \texttt{CircleBody} extends the parent \texttt{Body}; defining the geometry to be a circle and supplying some simplified methods such as the distance function. \texttt{FloodPlot} is a visualization class which makes nice flood plots to extend \texttt{Field.display}. Within \texttt{setup} we have:

\begin{verbatim}
  int n = (int)pow(2,6);                   // number of grid points
  size(400,400);                           // display window size
  Window view = new Window(n,n);

  body = new CircleBody(n/3,n/2,n/8,view); // define geom
  flow = new BDIM(n,n,1.5,body);           // define flow using BDIM
  flood = new FloodPlot(view);
  flood.setLegend("vorticity",-.75,.75);
\end{verbatim}
The first line sets the number of points in the domain. This needs to be a large power of two (Java uses the \texttt{pow} function for exponents) to allow sub-division in the multi-grid solver.\footnote{Multi-grid methods are especially well suited to object oriented programming, enabling a complete implementation in less than 90 lines of code.} Next the size of the pop-up display is defined in number of pixels. The third defines a \texttt{Window} object to map the numerical grid to a portion of the display. Any number of windows may be defined and used to plot different `views' of the simulation.

In the next block, the body is constructed. The arguments define the size of the circle, and its location in the simulation. This process is indefinitely extendable - any number of additional bodies of any size or shape can be defined to form a composite \texttt{Body} object. Note that the arguments are all given in terms of the number of grid cells, ie the diameter is $n/8=8$ meaning the circle is $8$ grid cells wide. In Lily Pad all distances are given in terms of the unit grid spacing since it is the only predetermined length scale. This has the pedagogical value of making all dimensions into statements of ``resolution''. Since $D$ is only 8, we can safely assume this simulation is under-resolved. If we increase the number of points, this will increase the resolution by the same factor, etc.

Next, the flow is constructed using the number of points, time step, and body object as arguments. The default \texttt{BDIM} object uses a unit background flow $\vec U = (1,0)$, but any velocity field can be defined and used as an initial condition. Since the distance scale is the grid spacing, the time step $\Delta t$ is equal to the CFL number. If $\Delta t$ is set to zero, adaptive time stepping is used with $\Delta t = (\max_{\vec x}|u|+3\nu)^{-1}$ to ensure explicit stability of the convective and diffusive terms. The kinematic viscosity $\nu$ is an optional argument to the \texttt{BDIM} constructor and its value needs to be set using the grid length scale. For example to match $Re_D=UD/\nu=1000$ the grid-based Reynolds number is $Re_g = U/\nu = Re_D/D = 125$ and $\nu=0.008$. Again, there is pedagogical value in this approach - $\nu$ is so small that the physical viscosity is likely to be drowned out by the ILES subgrid damping. Finally, the \texttt{FloodPlot} object is constructed by passing it a view, and then setting the color legend text and limits.

It is worth emphasizing that the flow solver has been set-up \textit{in four lines of code}. In standard CFD software, setting-up the fluid domain requires tremendous time and effort and entails the explicit or implicit specification of thousands of numerical parameters (the grid!). In Lily Pad, the fluid domain set-up is a non-issue and the only effective free numerical parameter is the `resolution'. \footnote{The kernel width $\epsilon$ is not a free parameter. Previous studies have shown that $\epsilon=2$ is optimal.}

The Processing \texttt{draw} method is looped indefinitely and is used to iterate the update and display calls:
\begin{verbatim}
  body.update();
  flow.update(body); flow.update2();
  flood.display(flow.u.vorticity());
  body.display();
\end{verbatim}
In this code, the body is updated, taking into account any changes input by the user. Alternatively, the body can be updated using prescribed motion or fluid-structure interaction equations. The updated body is used to update the conditions for the flow, and the flow is integrated in separate predictor and corrector steps. Next the flood plot is used to display the vorticity of the flow velocity\footnote{Simple chaining of commands is another advantage of the object oriented coding style.} and the body is displayed on top. The display commands have been optimized for speed, and the educational value of watching the flow develop  in real-time can hardly be overstated.

\section{Advanced Examples and Next Steps}

\begin{figure}[t!]
	\centering
	\subfloat[]{
		\includegraphics[trim=0cm 1cm 2cm 1cm, clip=true, width=0.45\textwidth]{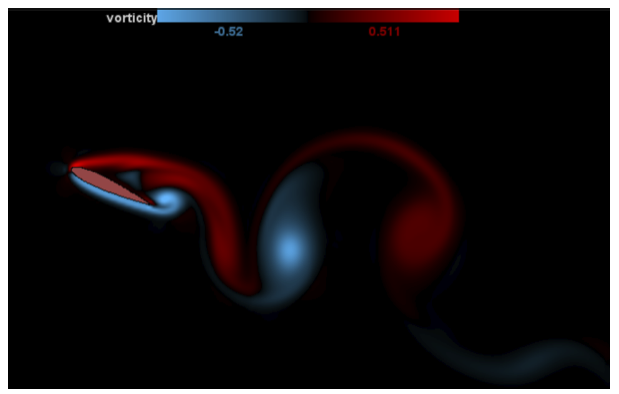}
		\label{fig:amanda}}
	\subfloat[]{
		\includegraphics[trim=0cm 0cm 3cm 0cm, clip=true, width=0.45\textwidth]{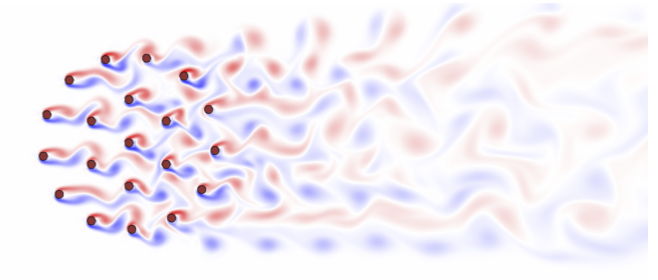}
		\label{fig:chen}}\\
	\subfloat[]{
		\includegraphics[trim=1cm 1cm 1cm 1cm, clip=true, width=0.45\textwidth]{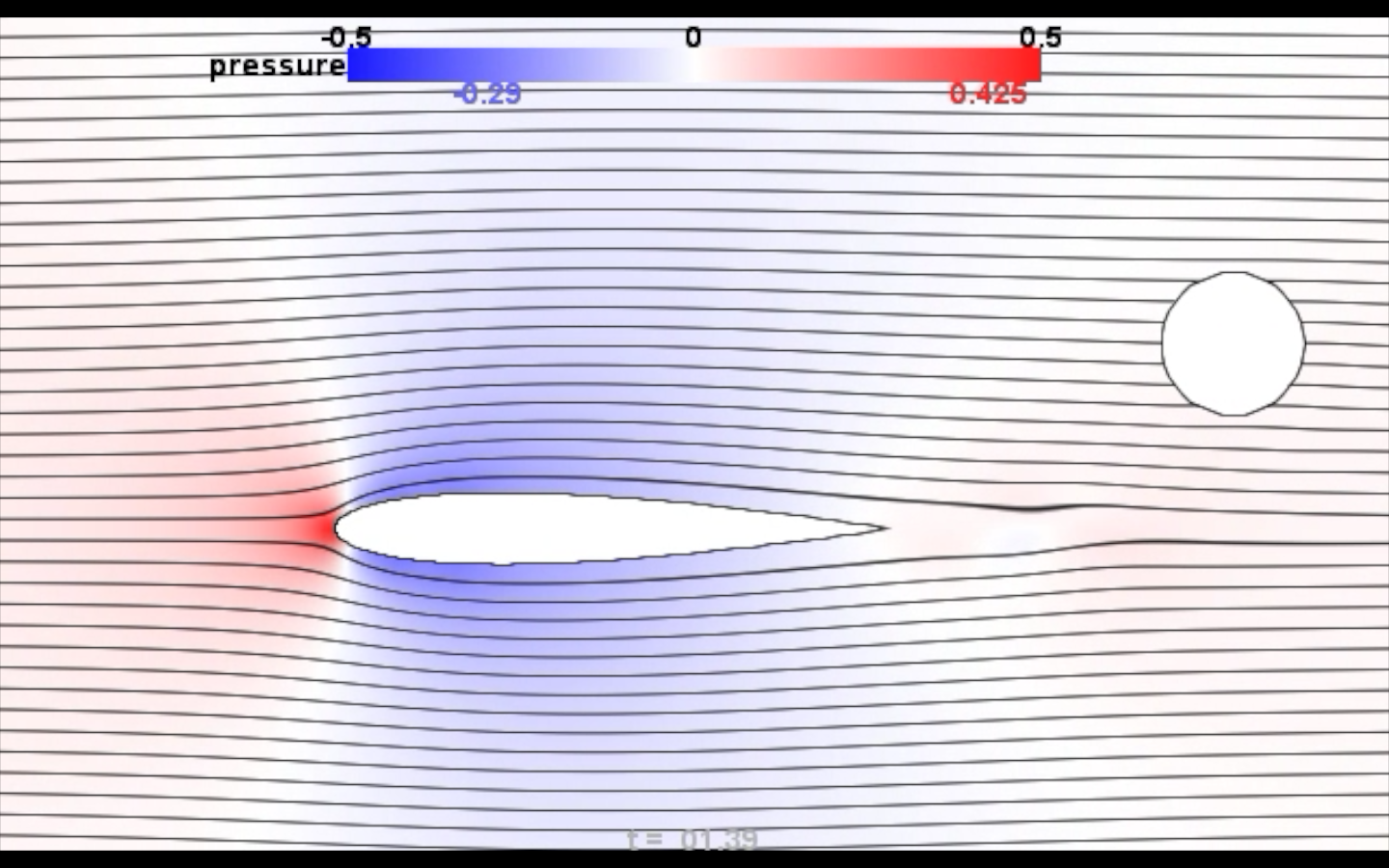}
		\label{fig:audrey}}
	\subfloat[]{
		\includegraphics[trim=3cm 1cm 5cm 1cm, clip=true, width=0.45\textwidth]{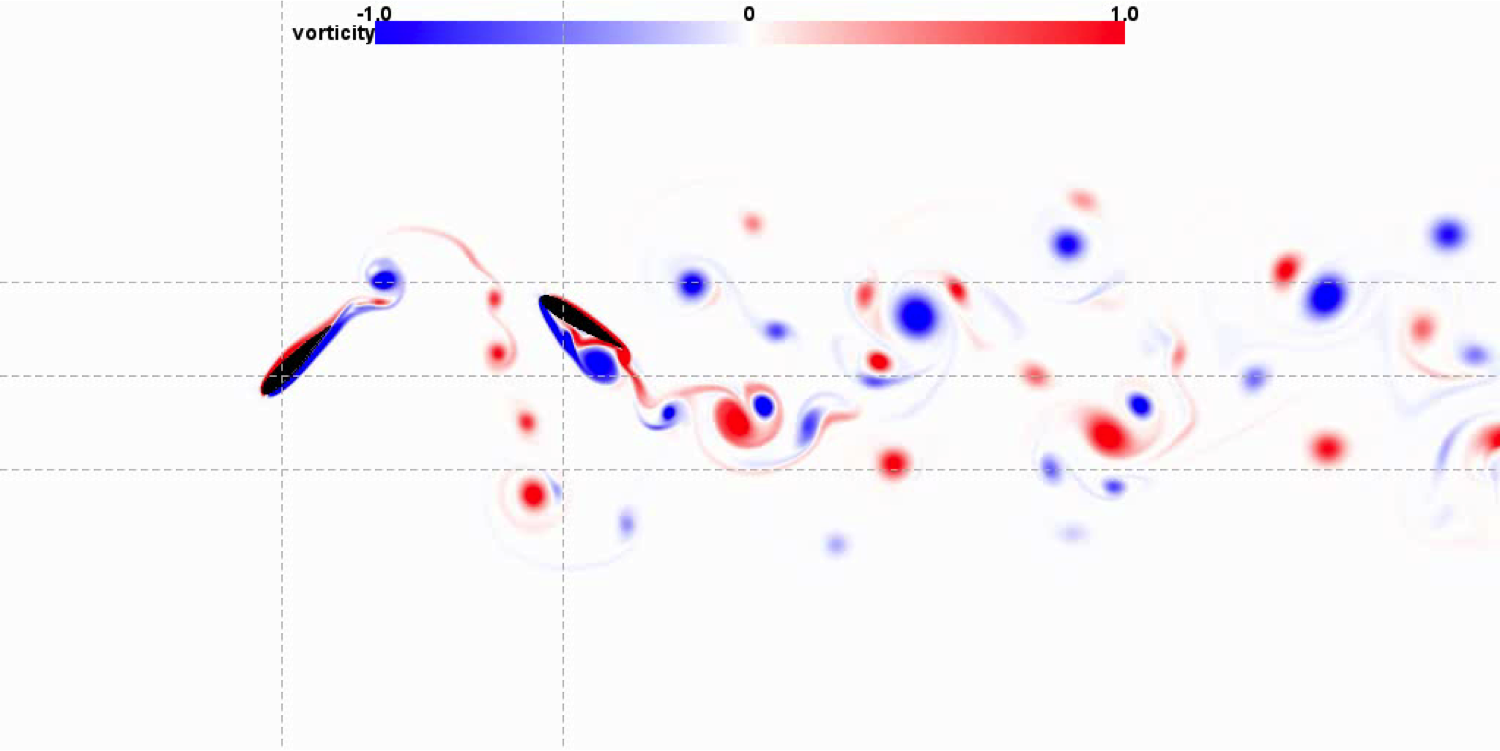}
		\label{fig:luke}} \\
	\subfloat[]{
		\includegraphics[width=0.45\textwidth]{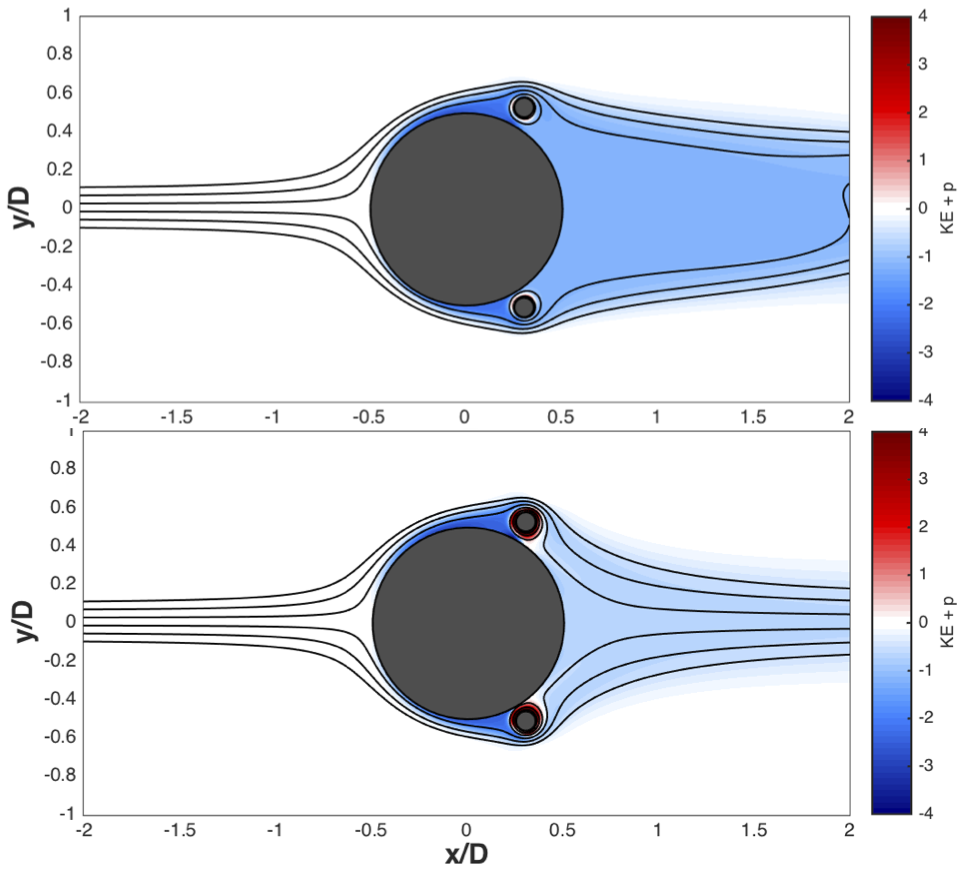}
		\label{fig:james}}
	\subfloat[]{
		\includegraphics[trim=1cm 2cm 1cm 15mm, clip=true, width=0.45\textwidth]{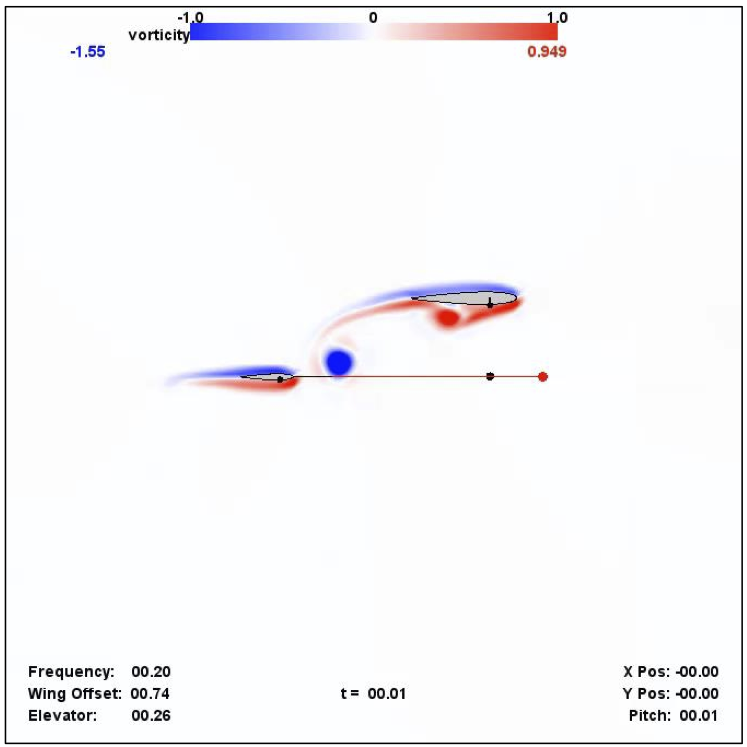}
		\label{fig:jacob}}
	\caption{Lily Pad example applications courtesy of (a) Amanda Persichetti, (b) Chen Yongxin, (c) Audrey Maertens, (d) Luke Muscutt, (e) James Schulmeister, and (f) Jacob Izraelevitz.}
	\label{fig:projects}
\end{figure}

This default example gives a basic introduction to the grammar used in a Lily Pad script, but one of the software's strengths is how easily the methods are extended to different flow types. The most basic extension is to change the body's shape and dynamics, such as in \cite{Weymouth2014JFM,Polet2015} or a flapping foil energy extractor (Fig~\ref{fig:amanda}). Increasing the number of bodies is also very simple, such as an array of 20 offshore risers (Fig~\ref{fig:chen}). Combining multiple bodies with independent dynamics extends the potential application: from blind fish swimming past obstacles (Fig~\ref{fig:audrey}) and drag reducing spinning cylinders (Fig~\ref{fig:james}) to tandem swimming flippers of a turtle (Fig~\ref{fig:jacob}) or even a Plesiosaur (Fig~\ref{fig:luke}). Extensions can also be made to the visualization and user interface, as seen on the Lily Pad website \cite{lilypad}.

However, there are still limitations in this approach. The fundamental drawback is being restricted to two-dimensional flow. A three-dimensional version, called Lotus, is in development but 3D simulations increase the computational cost by many orders of magnitude. Lotus attempts to compensate for this by implementing multiprocessor versions (MPI and CUDA) and fast programming and visualization languages (Fortran/Paraview) but is still much slower than real-time. Advances in 2D/3D models may help bridge the gap, at least in the initial design and simulation set-up stages. High Reynolds number simulations will also require modeling advances, particularly in the development of near-wall models appropriate for Immersed Boundary methods.

Nevertheless, two-dimensional and moderate Reynolds number simulations still have great value, especially in education. All the results in Fig~\ref{fig:projects} are student projects, and most were developed by students without formal training in CFD. Lily Pad's ability to quickly and easily go from concept to simulation has helped all of these students add depth to their studies. In addition, Lily Pad has been used to develop stand-alone demo applications which are excellent for education and outreach \cite{mainsail,mit-homepage}. As such, Lily Pad has at least partially achieved its goal of lowering the barrier to CFD, and is a first step towards more general real-time interactive solvers.

\bibliographystyle{plain}
\bibliography{../../complete}

\begin{thebibliography}{10}

\bibitem{lilypad}
Lily pad: Real-time two-dimensional fluid dynamics simulations in processing.
\newblock \url{http://github.com/weymouth/lily-pad}.
\newblock Accessed: 2015-09-29.

\bibitem{Aspden2008}
Stuart~Dalziel Andrew~Aspden, Nikos~Nikiforakis and John~B. Bell.
\newblock Analysis of implicit les methods.
\newblock {\em Communications in Applied Mathematics and Computational
  Science}, 3(1):103--126, 2008.

\bibitem{mainsail}
{CNN MainSail}.
\newblock How has foiling made boats much faster?
\newblock
  \url{http://edition.cnn.com/videos/tv/2015/03/11/spc-mainsail-design-special-b.cnn/video/playlists/intl-mainsail}.
\newblock Accessed: 2015-09-29.

\bibitem{Couder1986}
Y~Couder and C~Basdevant.
\newblock Experimental and numerical study of vortex couples in two-dimensional
  flows.
\newblock {\em Journal of Fluid Mechanics}, 173:225--251, 1986.

\bibitem{Groen2015}
Derek Groen, Xiaohu Guo, James~A Grogan, Ulf~D Schiller, and James~M Osborne.
\newblock Software development practices in academia: a case study comparison.
\newblock {\em arXiv preprint arXiv:1506.05272}, 2015.

\bibitem{Kiara2013}
Areti Kiara, Kelli Hendrickson, and Dick~KP Yue.
\newblock Sph for incompressible free-surface flows. part i: Error analysis of
  the basic assumptions.
\newblock {\em Computers \& Fluids}, 86:611--624, 2013.

\bibitem{Maertens2015}
Audrey~P Maertens and Gabriel~D Weymouth.
\newblock Accurate cartesian-grid simulations of near-body flows at
  intermediate reynolds numbers.
\newblock {\em Computer Methods in Applied Mechanics and Engineering},
  283:106--129, 2015.

\bibitem{Margolin2013}
LG~Margolin.
\newblock Finite scale theory: The role of the observer in classical fluid
  flow.
\newblock {\em Mechanics Research Communications}, 2013.

\bibitem{openfoam}
{OpenFOAM Foundation}.
\newblock Openfoam.
\newblock \url{http://www.openfoam.org}.
\newblock Accessed: 2015-09-29.

\bibitem{Peskin1972}
Charles~S. Peskin.
\newblock Flow patterns around heart valves: A numerical method.
\newblock {\em Journal of Computational Physics}, 10(2):252 -- 271, 1972.

\bibitem{Polet2015}
Delyle~T. Polet, David~E. Rival, and Gabriel~D. Weymouth.
\newblock Unteady dynamics of rapid perching manoeuvres.
\newblock {\em Journal of Fluid Mechanics}, 767(323--341), 2015.

\bibitem{processing}
{Processing Foundation}.
\newblock Processing.
\newblock \url{http://www.processing.org}.
\newblock Accessed: 2015-09-29.

\bibitem{r-project}
{R Core Team}.
\newblock R: A language and environment for statistical computing.
\newblock \url{http://www.R-project.org}.
\newblock Accessed: 2015-09-29.

\bibitem{Rouson2011}
Damian Rouson, Jim Xia, and Xiaofeng Xu.
\newblock {\em Scientific software design: the object-oriented way}.
\newblock Cambridge University Press, 2011.

\bibitem{Weymouth2014JFM}
G.~D. Weymouth.
\newblock Chaotic rotation of a towed elliptical cylinder.
\newblock {\em Journal of Fluid Mechanics}, 743:385--398, 3 2014.

\bibitem{mit-homepage}
Gabriel~D Weymouth.
\newblock Interactive research examples.
\newblock \url{http://web.mit.edu/weymouth/www}.
\newblock Accessed: 2015-09-29.

\bibitem{Weymouth2011JCP}
Gabriel~D. Weymouth and Dick K.-P. Yue.
\newblock Boundary data immersion method for cartesian-grid simulations of
  fluid-body interaction problems.
\newblock {\em Journal of Computational Physics}, 230(16):6233--6247, July
  2011.

\end{thebibliography}

\end{document}